\documentclass[
reprint, superscriptaddress,
 amsmath,amssymb,
 aps,
 pra,
]{revtex4-1}

\usepackage{graphicx}
\usepackage{overpic}
\usepackage{dcolumn}
\usepackage{bm}
\usepackage[hidelinks,colorlinks=true,linkcolor=blue,citecolor=blue]{hyperref}
\usepackage[mathlines]{lineno}
\usepackage{titlesec}
\usepackage{braket}
\usepackage{color}
\usepackage{float}

\begin{document}


\title{Two Ultracold Atoms  in a Quasi-Two-Dimensional Box  Confinement}

\author{Fan Yang}
\affiliation{Department of Physics, Renmin University of China, Beijing, 100872, P. R. China}
\affiliation{Key Laboratory of Quantum State Construction and Manipulation (Ministry of Education), Renmin University of China, Beijing, 100872, China}

\author{Ruijie Du}

\affiliation{Department of Physics, Renmin University of China, Beijing, 100872, P. R. China}
\affiliation{Key Laboratory of Quantum State Construction and Manipulation (Ministry of Education), Renmin University of China, Beijing, 100872, China}

\author{Ran Qi}
 \email{qiran@ruc.edu.cn}

\affiliation{Department of Physics, Renmin University of China, Beijing, 100872, P. R. China}
\affiliation{Key Laboratory of Quantum State Construction and Manipulation (Ministry of Education), Renmin University of China, Beijing, 100872, China}

 \author{Peng Zhang}%
 \email{pengzhang@ruc.edu.cn}

\affiliation{Department of Physics, Renmin University of China, Beijing, 100872, P. R. China}
\affiliation{Key Laboratory of Quantum State Construction and Manipulation (Ministry of Education), Renmin University of China, Beijing, 100872, China}

\date{\today}

\begin{abstract}

We investigate the scattering and two-body bound states of two ultracold atoms in a quasi-two-dimensional (quasi-2D) confinement, with the
confinement potential being an infinite square well (box potential)
 in the transverse ($z$-) direction, and the motion of the atoms in the $x$-$y$ plane being free. Specifically,
we calculate the effective 2D scattering length  and 2D effective range  of the low-energy scattering, as well as the energy   and the
transverse-excited-mode probability  of the bound states.
Comparing these results with those obtained
under a harmonic transverse confinement potential,
we find that  in most of the cases the 2D effective range for the box confinement is approximately 0.28 of the one for the harmonic confinement. Moreover, the transverse-excited-mode probability of the bound states for the box confinement is also much lower than the one for the harmonic confinement.
These results suggest that
the  transverse excitation
 in the  box confinement
  is notably  weaker than the one  in a harmonic confinement. Therefore,
  achieving quasi-2D ultracold gases well-described by pure-2D effective models, particularly those with 2D contact interaction, is more feasible through box confinement. Our results are helpful for the quantum simulation of 2D many-body physics with ultracold atoms, {\it e.g.}, the suppression of 2D effective range may lead to an enhancement of quantum anomaly in two-dimensional Fermi Gases. Additionally, our calculation method is applicable to  the two-body problems of  ultracold atoms in other types of quasi-2D confinements.

\end{abstract}

\maketitle


\section{\label{sec:level1}Introduction}

The combination of  Feshbach resonance and  optical lattices provide a very useful toolbox to realize low dimensional systems in ultra-cold quantum gases \cite{1D1,1D2,1D3,1D4,1D5,2D1,2D2}. In such setups, the motion of atoms along certain dimensions are effectively frozen out by the strong transverse confinement from optical lattices, while the motion along other dimensions remains free. As a result, the system can be described by a simple effective low-dimensional Hamiltonian,
and thus can be used to explore different kinds of strongly correlated many-body systems, such as Tonk and super-Tonk gas, Berezinskii-Kosterlitz-Thouless transition \cite{1D3,1D4,1D5,2D2}.

In the quasi-low-dimensional ultracold gases, the inter-atomic interaction can couple the transverse ground state and excited states,
and induce the following two effects: (1) During the scattering process, there are virtual transitions between the transverse ground and excited modes.
 (2) In the low-energy two-atom bound state, the atoms have non-zero probability to be in the transverse excited states. When these two effects are significant,
  both  the effective low-dimensional scattering length
and the effective low-dimensional effective range make non-negligible contributions to the scattering amplitude, and the inter-atomic interaction cannot be described by a simple low-dimensional contact potential. A more complicated interaction model, {\it e.g.}, a multi-channel model, should be applied. This brings complications to the quantum simulation of low-dimensional many-body physics. One important example is that a non zero 2D effective range leads to a significant reduction of the quantum anomaly in two-dimensional Fermi Gases \cite{Hu1,Liu1,Hu2}.

In most of the current experiments of quasi-low-dimensional ultracold gases, the transverse confinement potentials are nearly harmonic. As pointed out in \cite{Duan1}, in these systems the above effects caused by the interaction-induced coupling between the transverse ground  and excited modes are usually non-negligible, and it is not efficient to suppress these effects simply by increasing the strength of trapping potential.
On the other hand, in recent years the box trap, {\it i.e.}, the trap with confinement potential being an infinite square well (box potential), has been realized in more and more experiments, and provided a whole new approach to confine atoms, including in quasi-2D confinement \cite{box1, box2, box3, box4}.
Then it is nature to ask:
\begin{itemize}
\item[] {\it If the transverse confinement of a
quasi-low-dimensional ultracold gas is realized via a box confinement potential, rather than a harmonic one, can the above interaction-induced transverse transition effects be suppressed?}
\end{itemize}

Notice that there are the following two important differences between the box and harmonic transverse confinements, which are related to the above question:
\vspace{0.15cm}

(A) The level spacing of box trap is not uniformly distributed but increasing rapidly for higher excited states.     Intuitively speaking,
  the transverse excitations
  should be suppressed by these large energy gaps.
\vspace{0.15cm}

(B) The
   box confinement potential couples the two-atom relative motion
 to the center-of-mass (CoM) motion. As a result, in this confinement the transverse ground state is coupled to more transverse excited states. This fact may increase the transverse excitation probability.
 \vspace{0.15cm}

\noindent Since there are both of these two competing mechanisms,
  one can answer the above question only by quantitative calculations.

  \begin{figure}[tbp]
  \centering
\begin{minipage}[b]{1\linewidth}
  \begin{overpic}[width=6.5cm]{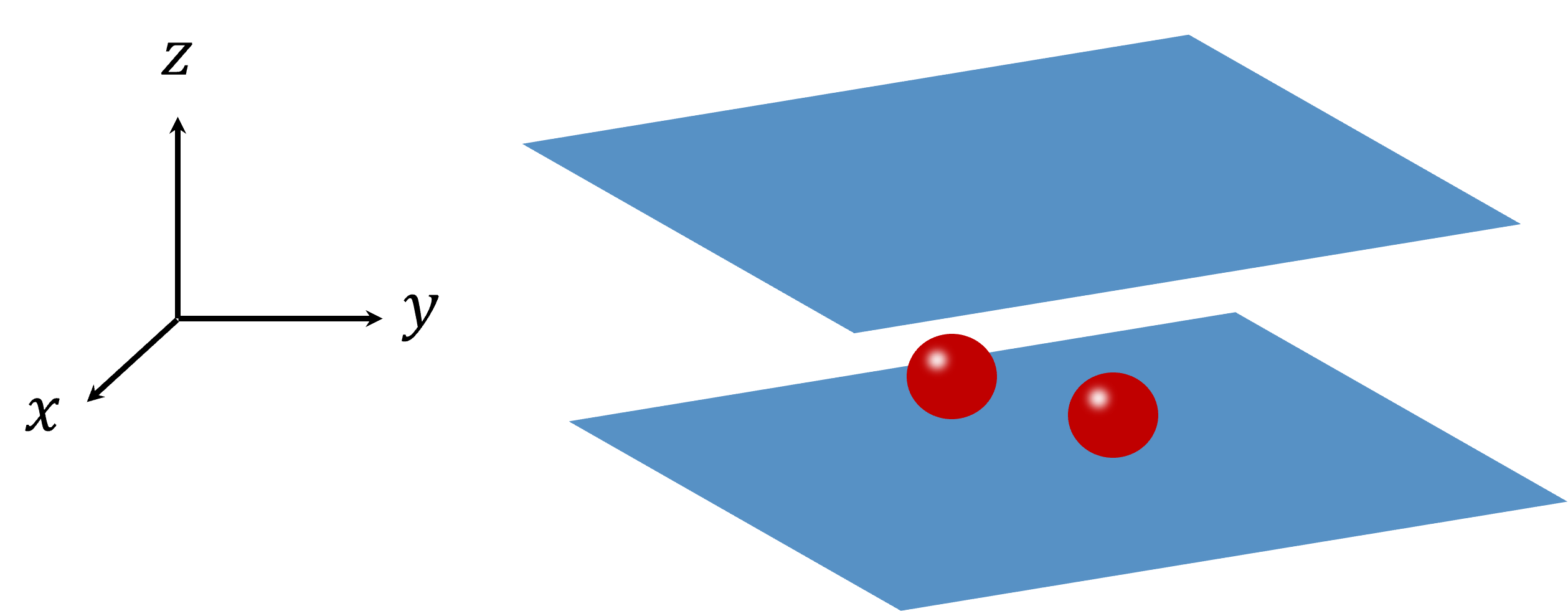}
    \put(0,36) {(a)}
  \end{overpic}
  \vspace{0.8cm}
\end{minipage}
\begin{minipage}[b]{1\linewidth}
  \begin{overpic}[width=3.5cm]{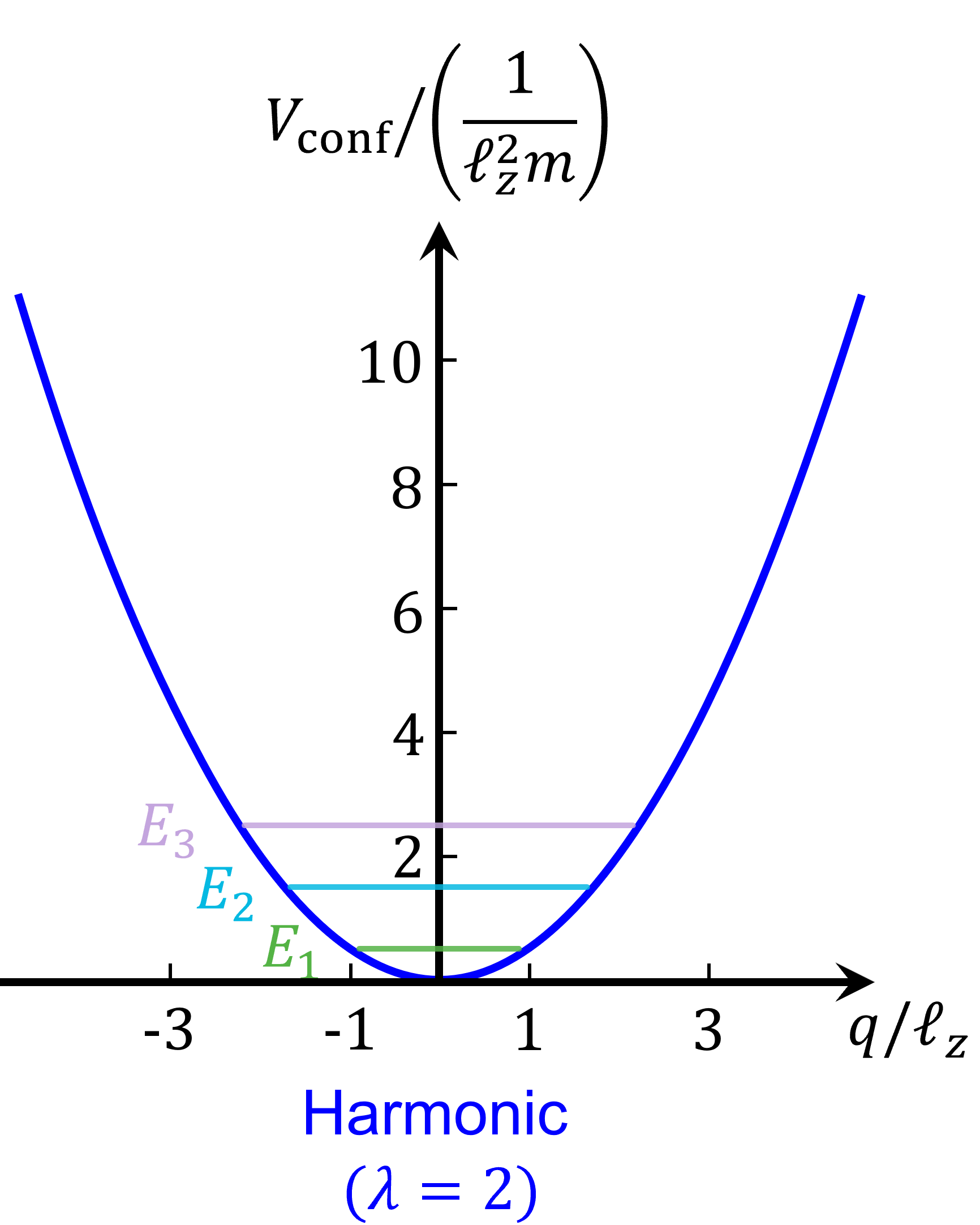}
    \put(0,90) {(b)}
  \end{overpic}
  \begin{overpic}[width=3.5cm]{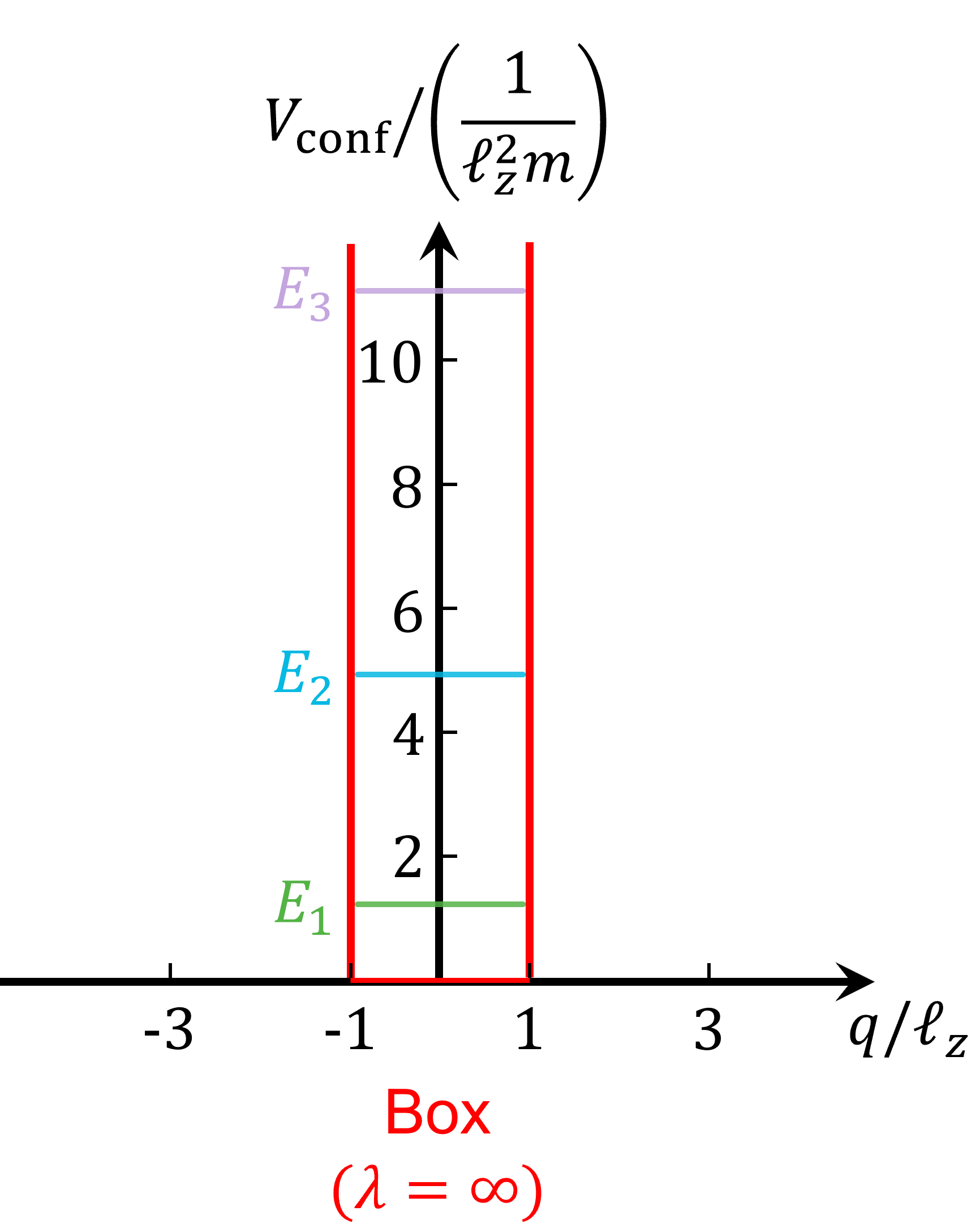}
    \put(0,90) {(c)}
  \end{overpic}
\end{minipage}
\caption{\textbf{(a)}: Schematic diagram of two interacting ultracold cold atoms in a quasi-2D  setup, which is realized via a strong  confinement potential  is along the transverse ($z$-) direction. \textbf{(b, c)}: The harmonic ($\lambda=2$) and box ($\lambda=\infty$) transverse confinement potential $V_{\rm conf}$,  which are given by Eq.~(\ref{eq.curve}). Three lowest eigen-energies of each potential are marked with horizontal lines.}
\label{scheme}
\end{figure}

In this work we consider this  problem for quasi-two-dimensional (quasi-2D) ultracold gases (Fig.~\ref{scheme}(a)). Specifically, we calculate the scattering length $a_{\rm 2D}$ and effective range $R_s$ for the effective 2D scattering of two ultracold atoms in a quasi-2D
setup realized via
 box confinement potential a (Fig.~\ref{scheme}(c)), as well as  the energy $E_b$ and transverse-excited-state probability $P_{\rm ex}$ of the bound states of these two atoms. We further compare these results with the ones of a  quasi-2D  harmonic confinement (Fig.~\ref{scheme}(b)), and show that in most of the cases the value of  $R_s$ for the box confinement is approximately 0.28 of the one for the harmonic confinement, and
the value of $P_{\rm ex}$ for the box confinement is also significantly lower than the one for the harmonic confinement. As discussed above, these results yield that the box transverse confinement potential can suppress the interaction-induced transverse excitation effects. Our results are helpful for the studies of quantum simulation with quasi-low-dimensional ultracold gases. Additionally, our calculation approach can also be used for the
quasi-low-dimensional two-body problems with other types of transverse confinement potential.

The remainder of this paper is organized as follows. In Sec.~\ref{sec:level2}
we introduce the our system and the to-be-calculated quantities in detail.
The details of our calculation approach are shown in the appendixes.
In Sec.~\ref{res}
we show the obtained results, and compare the ones for box and harmonic confinements. A summary of this work is given in Sec.~\ref{summary}.


\section{Two atoms in a Quasi-2D Confinement}
\label{sec:level2}
We consider two equal-mass ultracold atoms
 in a quasi-2D setup realized via a transverse confinement potential which is applied in the $z$-direction
 (Fig.~\ref{scheme}(a)). For this system,  the CoM motion of these two atoms in the $x$-$y$ plane can be separated out. As a result, we can only consider the  relative motion in all the three directions, as well as the
  CoM motion along the $z$-direction.
  The Hamiltonian of our system is given by ($\hbar=1$):
 \begin{eqnarray}
    \hat{H} &=& -\frac{1}{2\mu}\nabla_{\bm{r}}^2  -\frac{1}{2M}\frac{\partial^2}{\partial Z^2}+\sum_{j=1,2}V_{\rm conf}(z_j)+U_{\rm HY}(\bm{r}),\nonumber\\
    \label{h}
\end{eqnarray}
where $M=2m$ and $\mu=m/2$ are the total and reduced mass of these two atoms, respectively,
with $m$ being the single-atom mass,
$z_j$ ($j=1,2$) is the coordinate of the atom  $j$ in the $z$-direction,
$Z=(z_1+z_2)/2$ is the center-of-mass (CoM) coordinate  in the  $z$-direction.
Additionally,   $V_{\rm conf}$ is the confinement potential, and
is assumed to be expressed as
\begin{equation}
  V_{\mathrm{conf}}(q) = \frac{1}{2m\ell_z^2}\left(\frac{q}{\ell_z}\right)^\lambda,
  \label{eq.curve}
\end{equation}
where $l_z$ is the characteristic length of the confinement, and the parameter $\lambda$ takes the values $2$ or $\infty$. Explicitly, when $\lambda= 2$ the confinement potential is  harmonic with angular frequency $\omega=1/(m\ell_z^2)$ (Fig.~\ref{scheme}(b)), while when $\lambda=\infty$ the confinement potential is an infinite square well with width $2\ell_z$ (Fig.~\ref{scheme}(c)).
Moreover, in Eq.~(\ref{h})
 $U_{\rm HY}(\bm{r}) $ is the Huang-Yang pseudopotential \cite{LHY}:
 \begin{eqnarray}
 U_{\rm HY}(\bm{r}) &\equiv&\frac{2\pi a_{\rm{3D}}}{\mu}\delta(\bm{r})  \frac{\partial}{\partial r} (r\cdot),\label{uhy}
\end{eqnarray}
which describes the
the inter-atomic interaction, with ${\bm r}=(x,y,z)$ being the relative coordinate of the two atoms,
$r=|{\bm r}|$, and
 $a_{\rm{3D}}$ being the $s$-wave scattering length of the two atoms in the 3D free space.

For the convenience of our discussion, we further define $\{E_n,\phi_n(q)\}$ ($n=1,2,3,...$) as the eigen energies and normalized eigen wave functions of the one-dimensional (1D) single-atom Schr\"odinger equation with potential $V_{\mathrm{conf}}(q)$, {\it i.e.},
\begin{eqnarray}
 \left[-\frac{1}{2m}\frac{\partial^2}{\partial q^2} + V_{\mathrm{conf}}(q)\right]\phi_n(q) = E_n\phi_n(q),\nonumber\\
(n=1,2,...),
\end{eqnarray}
and $E_1<E_2<E_3<...$.

We consider the scattering of these two atoms, which are all in the transverse ground state $\phi_1$. Explicitly, the incident state of this scattering process is
\begin{eqnarray}
\Psi_{\bm k}^{\rm(in)}({\bm \rho},z_1,z_2)=\frac{1}{2\pi}e^{i{\bm k}\cdot{\bm \rho}}\phi_1(z_1)\phi_1(z_2),\label{incident}
\end{eqnarray}
with ${\bm \rho}=(x,y)$, and ${\bm k}$ being the incident momentum in the $x$-$y$ plane. Here we focus on the low-energy cases with the relative kinetic energy $k^2/(2\mu)\ll E_2-E_1$. In these cases, the scattering wave function
satisfies the outgoing boundary condition, {\it i.e.},
\begin{eqnarray}
\lim_{\rho \rightarrow \infty}\!\!\Psi_{\bm k}^{(+)}\!\!\left({\bm \rho},\! z_1, \! z_2\right)\!=\!\frac{1}{2\pi}\!\!\left[e^{i{\bm k}\cdot{\bm \rho}}\!+\!f_{\rm 2D}(k)\frac{e^{ik\rho}}{\sqrt{\rho}}\right]\!\!\phi_1(z_1)\phi_1(z_2),\nonumber\\
\end{eqnarray}
with $k=|{\bm k}|$,  $\rho=|{\bm \rho}|$.
Here $f_{\rm 2D}(k)$ is proportional to the effective 2D scattering amplitude, and can be expressed as
\begin{eqnarray}
f_{\rm 2D}(k)\approx \sqrt{\frac{\pi}{2k}}e^{i\pi/4}\frac{-1}{i\frac{\pi}2-\gamma-\ln\left(k\frac{a_{\rm 2D}}{2}\right)
-\frac{1}{2}R_sk^2}
,\label{f2D}
\end{eqnarray}
in the low-energy limit, where $\gamma\approx 0.577$ is the Euler's constant.
In Eq.~(\ref{f2D}),
 $a_{\rm 2D}$
and $R_s$ are the effective 2D scattering length and effective range of our quasi-2D system, respectively \cite{Hu1, ftn2}, and $R_s<0$.
Both $a_{\rm 2D}$ and $R_s$
 are functions of the 3D scattering length $a_{\rm 2D}$ and  the confinement parameters $\lambda$ and $\ell_z$, {\it i.e.},
\begin{eqnarray}
a_{\rm 2D}&=&a_{\rm 2D}\left(a_{\rm 3D},\ell_z,\lambda\right);\\
R_{s}&=&R_{s}\left(a_{\rm 3D},\ell_z,\lambda\right).
\end{eqnarray}

Furthermore, the contribution of the 2D effective range $R_s$
to  $f_{\rm 2D}(k)$ can be ignored under the condition \cite{ftn3},
\begin{eqnarray}
R_sk^2\ll 1. \label{con}
\end{eqnarray}
In this case, the inter-atomic interaction effect is determined only by a single parameter, {\it i.e.}, the 2D scattering length $a_{\rm 2D}$, and thus the system can be described by a pure-2D model with a simple {\it 2D single-channel contact interaction} \cite{Petrov,pure2d1,pure2d2,pure2d3,pure2d4,2Dexp3}, with the intensity being a function of $a_{\rm 2D}$.
One of the most important phenomena in 2D quantum gases is the emergence of quantum anomaly, which manifests itself in the deviation of the breathing mode frequency $\omega_B$ from the scale invariant value $2\omega_0$ in 2D harmonically trapped quantum gases \cite{pure2d1, anomaly1, anomaly3, anomaly4, anomaly5}.
Nevertheless, as mentioned in Sec.~\ref{sec:level1},
if the condition (\ref{con}) is not satisfied, the inter-atomic interaction
should be described by a 2D finite-range or two-channel interaction, which is rather complicated \cite{Duan1,Duan2,Liu1,Hu2,Liu2}.  Furthermore, this 2D finite-range effect may significantly reduce the deviation $|\omega_B-2\omega_0|$ which makes experimental confirmation of 2D quantum anomaly rather difficult \cite{Hu1,Liu1,Hu2}.

In most of the current experiments of quasi-2D ultracold gases, the  confinement potential is harmonic ($\lambda=2$).
For this system,
due to the interaction-induced virtual transverse excitations in the scattering process,
the effective range is as large as $R_s=-\ell_z^2\ln 2\approx -0.69\ell_z^2$ \cite{Hu1}, which makes  the condition (\ref{con}) being not satisfied in many experiments \cite{2Dexp1,2Dexp2,2Dexp3}.


Another related problem is the energy and transverse excitation of the two-body bound state. In our system, there is at least one two-body bound state. The  energy $E_b$ of a bound-state  can be expressed as
\begin{eqnarray}
E_b=2E_1-E_{\rm binding},\label{ebb}
\end{eqnarray}
where $2E_1$ is the threshold energy of the free motion of these two atoms, and $E_{\rm binding}>0$ is the binding energy. Furthermore, similar as the scattering states discussed above, the normalized wave function $\Phi_{b}\left({\bm \rho}, z_1, z_2\right)$ of a bound state includes the components of both the
ground and excited transverse states:
\begin{eqnarray}
\Phi_{b}\left({\bm \rho}, z_1, z_2\right)=\sum_{j,s=1}^{+\infty}\phi_b^{(j,s)}({\bm \rho})\phi_j(z_1)\phi_s(z_2).
\end{eqnarray}
Therefore, the probability of the bound state in the  transverse excited states is:
\begin{eqnarray}
P_{\rm ex}
=1-\int d{\bm \rho} \vert\phi_b^{(1,1)}({\bm \rho})\vert ^2.
\end{eqnarray}

If the transverse excitation probability $P_{\rm ex}$ and
 the binding energy $E_{\rm binding}$ of a bound state
 satisfy the conditons
\begin{eqnarray}
P_{\rm ex}&\ll& 1;\label{con2a}\\
E_{\rm binding}&\approx&
\frac{2}{\mu a_{\rm 2D}^2}e^{-2\gamma}\equiv E_u,
\label{con2b}
\end{eqnarray}
where $a_{\rm 2D}$ is the effective 2D scattering length of this system, then one can still use the simple
 pure-2D contact interaction model to describe the
 physics related to this bound state.
Otherwise, a more complicated interaction model is required.

In this work we investigate if  the conditions (\ref{con}, \ref{con2a}, \ref{con2b})
are easier to be satisfied when the harmonic transverse confinement potential of a quasi-2D gas
is replaced with a box confinement potential. To this end, we calculate
the parameters $(a_{\rm 2D}, R_s, E_{\rm binding}, P_{\rm ex})$ for the systems with a box transverse confinement ($\lambda=\infty$). Our method for these calculations is shown in the appendixes. In the next section, we compare the results we obtained with the ones for harmonic transverse confinement ($\lambda=2$) and perform a detailed analysis.

%
%

%
%
%

%
%
%
%

\section{Results}
\label{res}

In this section, we illustrate the parameters $(a_{\rm 2D}, R_s, E_{\rm binding}, P_{\rm ex})$ of the quasi-2D system with a box confinement potential, which is obtained via the methods in the appendixes,
and compare them with the ones of a harmonic confinement potential ($\lambda=2$).
We will
show that in most of the cases, it is easier to realize the conditions (\ref{con}, \ref{con2a}, \ref{con2b}) with a
  box confinement potential.

  \begin{figure}[tbp]
  \centering
  \begin{overpic}[width=7cm]{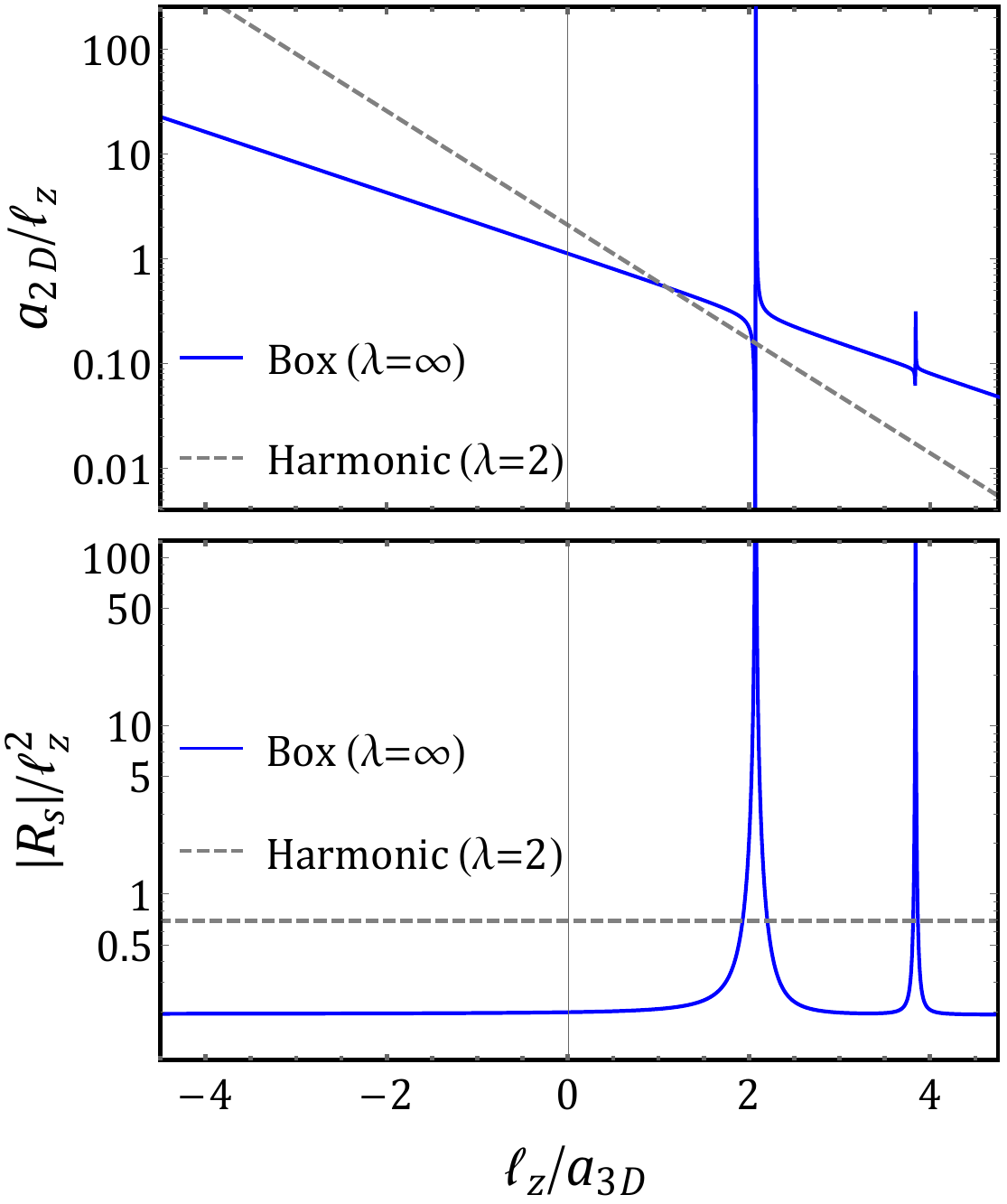}
    \put(1,97){(a)}
    \put(1,52.2){(b)}
  \end{overpic}
  \caption{The quasi-2D scattering length $a_{\rm{2D}}$ $\textbf{(a)}$,  and
  the absolute value $|R_{s}|$ of the
  quasi-2D effective range  \textbf{(b)}. Since $R_s$ is always negative, we have
  $|R_{s}|=-R_s$.
  Here we show the results for both the box confinement (blue solid lines)
 and the harmonic confinement (gray dashed lines). For the harmonic confinements, $a_{2D}$ and $R_s$ are given by Ref\cite{Petrov} and Ref\cite{Hu1}, respectively.
  }\label{Fig1}
\end{figure}

\subsection{Scattering Parameters $a_{\rm 2D}$ and $R_s$.}

  In Fig.(\ref{Fig1}) we illustrate the 2D scattering length $a_{\rm 2D}$ and effective range $R_s$ for both the systems with either a box or a harmonic transverse confinement. It is shown that  the behaviors of the $a_{\rm 2D}$ and $R_s$
 for these two confinements have the following two significant difference:

 First, as shown in Fig.~\ref{Fig1}(a), in the the box confinement we have
$a_{\rm 2D}=\infty$  when $a_{\rm 3D}\rightarrow 0^-$,
$a_{\rm 3D}\approx 0.48\ell_z$ and $a_{\rm 3D}\approx  0.26\ell_z$. There may be more resonances which are too narrow to be resolved by our calculation.
In contrast, for the harmonic confinement  $a_{\rm 2D}$ diverges only when $a_{\rm 3D}\rightarrow 0^-$.
Thus, there are at least two more 2D resonances   for the box confinement then for the harmonic confinement.
This difference is due to the fact that the box confinement potential can couple the CoM and relative motion in the $z$-direction, and induces more transverse excited states (closed channels) to be coupled to the incident state of the scattering. This effect also occurs in other  ultracold gases with the CoM and relative motion coupled by the confinement potential\cite{CIR,CIR1,CIR2,CIR3,mix1,mix2,mix3,mix4,mix5}

Second, as shown in Fig.~\ref{Fig1}(b), the effective range $R_s$ for the box confinement is {\it lower} than the one of the harmonic confinement, except in the region of the new 2D resonances with finite $a_{\rm 3D}$. Specifically, for the box confinement, we have $R_s\approx -0.19\ell_z^2$ in the regions out of  these new resonances, which is approximately 0.28 of the one for the harmonic confinement.
According to the analysis in the above sections, these results yield that in these regions the transverse excitation is eventually suppressed by the box confinement potential, and it is more easier to realize the condition (\ref{con}) with this confinement.
Additionally, Fig.~\ref{Fig1}(b) also shows that around the new resonances induced by the box confinements the effective range $R_s$ is significantly enhanced. Thus, the condition (\ref{con}) is difficult to be satisfied in these resonance regions.

\begin{figure}[tbp]
  \centering
  \begin{overpic}[width=7cm]{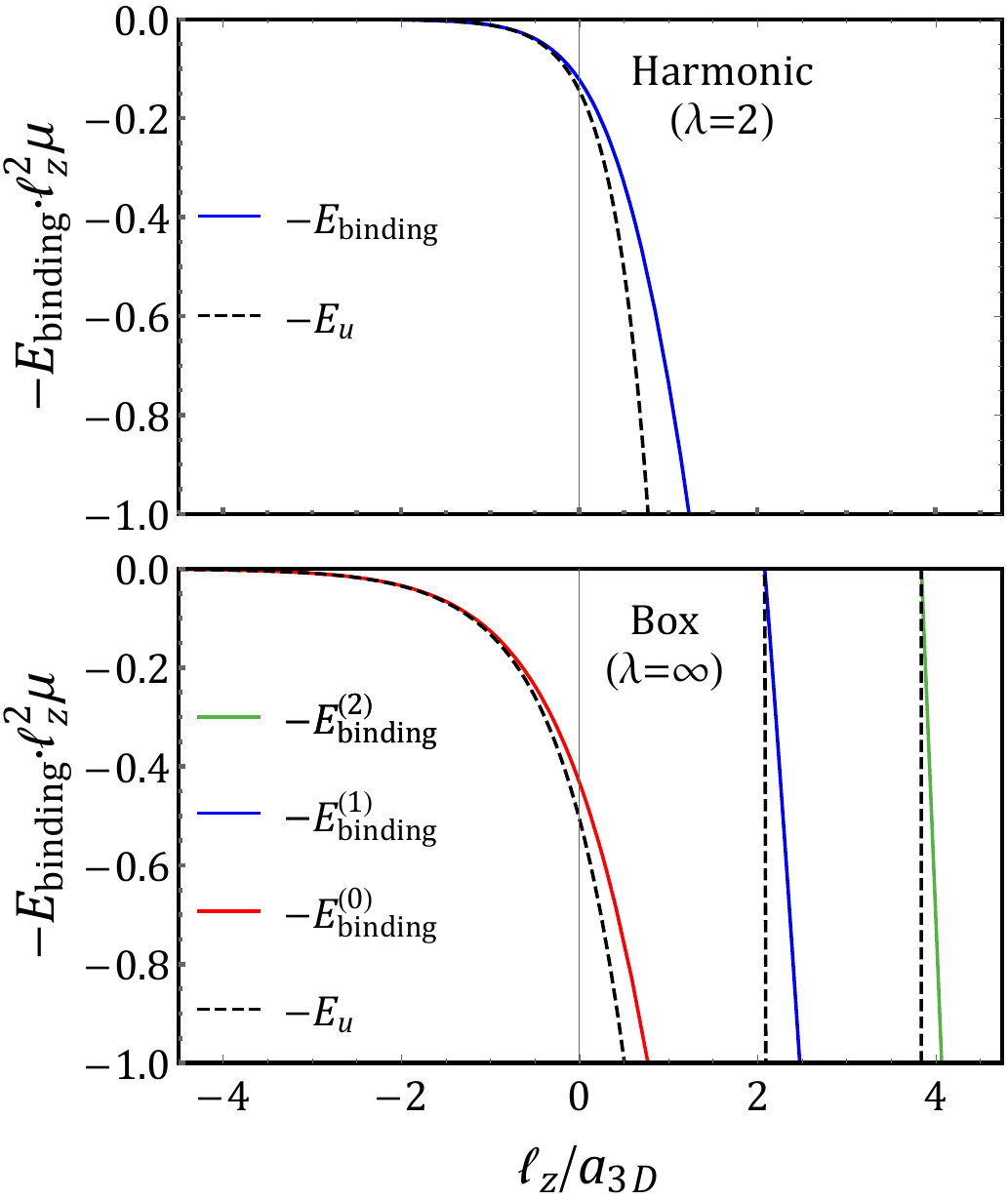}
    \put(1,97){(a)}
    \put(1,51){(b)}
  \end{overpic}
  \caption{ The negative of binding energy $-E_{\rm binding}$ (solid lines) and the energy $-E_u$ (dashed lines)
of the bound states of two atoms in a quasi-2D setup, with $E_u$ being defined in Eq.~(\ref{con2b}). Here we show the results for
 harmonic \textbf{(a)} and box \textbf{(b)} confinements.
}
  \label{Fig2}
\end{figure}

\subsection{Bound-State Parameters $E_{\rm binding}$ and $P_{\rm ex}$.}
\label{sec:level3}

In Fig(\ref{Fig2}), we show the negative of binding energy ({\it i.e.}, $-E_{\rm binding}$) for the bound states  of two atoms in a quasi-2D system with  harmonic and box confinements.
It is shown that for the harmonic confinement there is always one bound state
with the binding energy being zero at the 2D resonance $a_{\rm 3D}\rightarrow 0^-$. In contrast,
for the box confinement, one new bound state appears when the system cross each 2D resonance point.
Here we denote  $E_{\rm binding}^{(j)}$ ($j=0,1,2,...$) as the binding energy of the  $j$-th bound state, which satisfy
$E_{\rm binding}^{(0)}=0$ at $a_{\rm 3D}\rightarrow 0^-$,
$E_{\rm binding}^{(1)}=0$ at $a_{\rm 3D}\approx 0.48\ell_z$, $E_{\rm binding}^{(2)}=0$ at $a_{\rm 3D}\approx 0.26\ell_z$. Therefore, for harmonic confinement there is only the 0th bound state. Moreover, in Fig(\ref{Fig2}) the energy $-E_u$ corresponding to these bound states, which is defined in Eq.~(\ref{con2b}), is also illustrated. Fig(\ref{Fig2}) clearly shows that for the 0th bound state,  the condition (\ref{con2b}) ({\it i.e.}, $E_{\rm binding}\approx E_u$) is satisfied in a broader region of $E_{\rm binding}$ for the box confinement than for the harmonic confinement.

\begin{figure}[tbp]
  \centering
  \begin{overpic}[width=7cm]{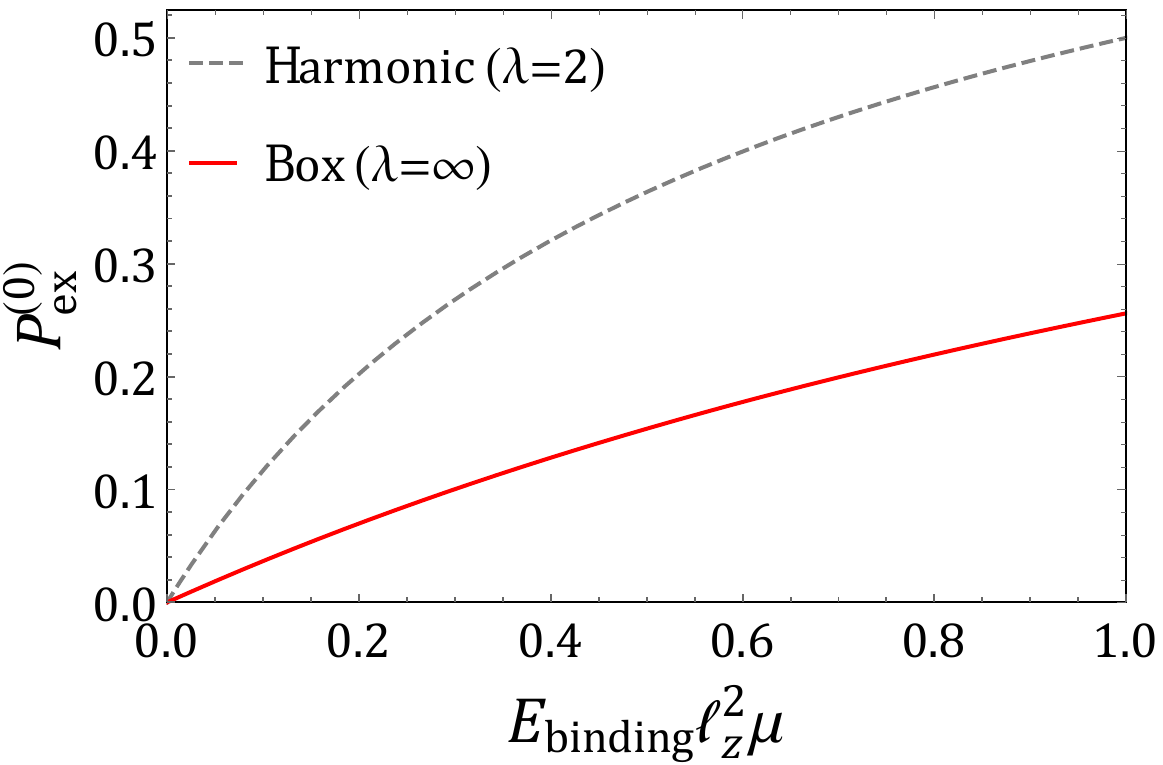}
    \put(1,62.2){(a)}
  \end{overpic}\vspace{0.02cm}
  \begin{overpic}[width=7cm]{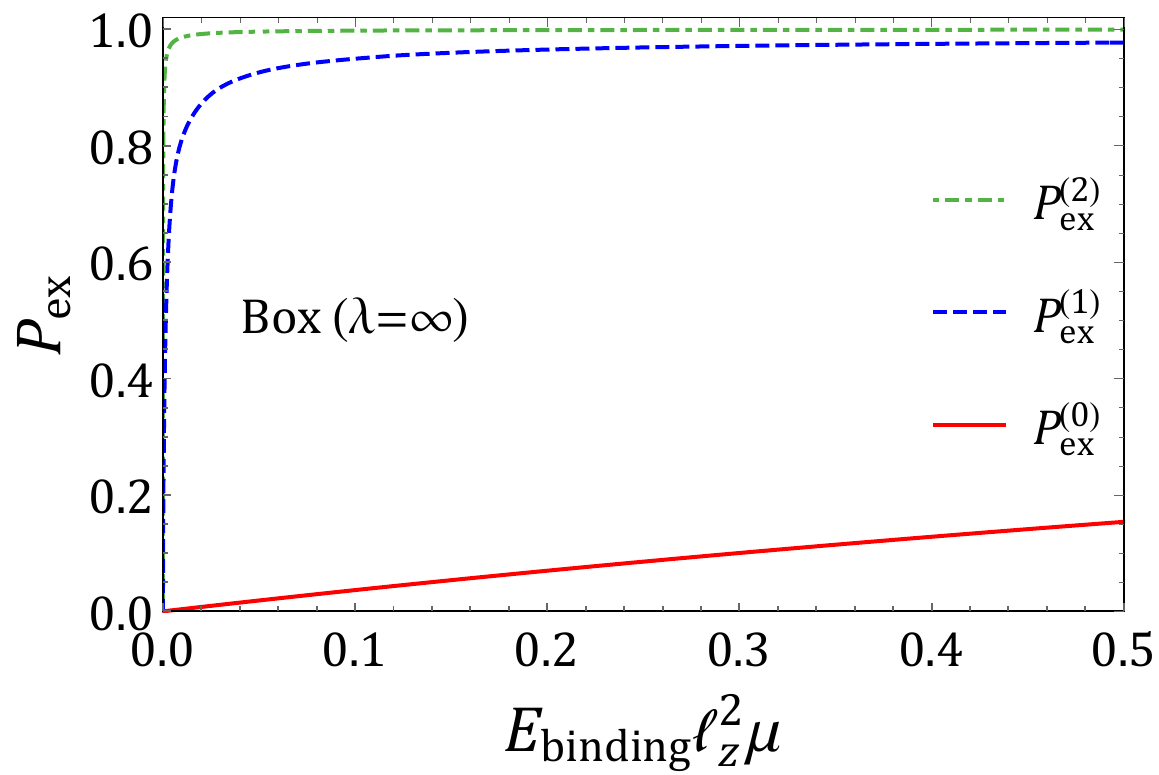}
    \put(1,63){(b)}
  \end{overpic}
  \caption{{\bf (a)}: The transverse excitation probability $P_{\rm ex}^{(0)}$ of the 0th bound state of two atoms in a quasi-2D setup with a box (red solid line) and harmonic (gray dashed line) confinement. {\bf (b)}: The transverse  excitation probability $P_{\rm ex}^{(0,1,2)}$ of the bound states for the cases with a box confinement.
  \label{fig.34}}
\end{figure}

Moreover, in Fig~\ref{fig.34} we show the transverse excitation probability $P_{\rm ex}^{(j)}$ for the $j$-th bound state ($j=0,1,2,...$), as functions of the energy $E_b^{(j)}$.
It is shown that for all the cases $P_{\rm ex}$ decreases with  $E_{\rm binding}$, and
$P_{\rm ex}\rightarrow 0$ in the limit $E_{\rm binding}\rightarrow 0$. Therefore, the condition (\ref{con2a}) can be realized in the low-energy region with the binding energy $E_{\rm binding}$ being low enough. Furthermore, as shown in Fig~\ref{fig.34}(a), the transverse excitation probability of the 0th bound state of the system with box confinement is much lower than the one with harmonic confinement. Thus,
the condition (\ref{con2a}) can be satisfied in a broader
region of
$E_{\rm binding}$
for the box confinement than for the harmonic confinement. This result is consistent with the above ones on the parameter region for the condition (\ref{con2b}). Additionally, as shown in Fig~\ref{fig.34}(b), the transverse excitation probability of the 1st and 2nd bound state is larger than the one of the 0th bound state, because the corresponding resonances are narrower. Thus, it is more difficult to realize the condition (\ref{con2a}) for these bound states.

\subsection{Conclusion}

Our aforementioned results indicate that it is easier to realize the conditions (\ref{con}, \ref{con2a}, \ref{con2b}) in a quasi-2D gas with a   box confinement than in the system with a harmonic confinement. This yields that the interaction-induced transverse excitation is eventually suppressed by the  box confinement. Thus, in this confinement the mechanism (A) mentioned in Sec.~\ref{sec:level1} prevails the mechanism (B).

\section{\label{summary}Summary}

In this paper, we first provide a general analysis of quasi-2D two-body scattering properties for two-particles confined in arbitrary external potentials along z-direction. In particular, we calculate scattering amplitude as well as bound state parameters for the case of a square well box confinement. We find that, for this box confinement, both the effective range and transverse excitation probability are significantly reduced compared to those of harmonic confinement with the same confining length. These results suggest that quantum gases in box confinement may be more accurately described by a single channel pure 2D model. The suppression of effective range may also lead to an enhancement of 2D quantum anomaly in Fermi gases confined by box potential in the transverse direction.

\begin{acknowledgments}

The authors thank Kuiyi Gao  for fruitful discussions.
This work was supported by the National Key Research and Development Program of China (Grant  No. 2022YFA1405301 (R.Q.), No. 2018YFA0306502 (R.Q.), and Grant  No. 2022YFA1405300 (P. Z.)),
NSAF Grant No. U1930201(P.Z.),
the  National Natural Science Foundation of China (Grant No. 12022405 (R.Q.) and No. 11774426 (R.Q.)), and
the Beijing Natural Science Foundation (Grant No. Z180013 (R.Q.)).

\end{acknowledgments}


\begin{widetext}
\appendix

\section{Calculation of Quasi-2D Scattering Length and Effective Range}

\subsection{Summary of the Main Steps}

In this appendix we derive and show our method for the calculation of quasi-2D scattering length $a_{\rm 2D}$ and effective range $R_s$. For the convenience of the readers, we first summarize the main steps of the our method in this subsection. The derivation of these steps would be given in the following sections.

We calculate $a_{\rm 2D}$ and $R_s$ via the following steps:

{\bf Step 1}: Numerically solving the integral equations (\ref{ne1}, \ref{ne2}), and
derive the functions $\eta^{(0)}(Z)$ and $\eta^{(2)}(Z)$. The definitions of the terms in Eqs.~(\ref{ne1}, \ref{ne2}) in the paragraph below these two equations.

{\bf Step 2}: Substituting the functions $\eta^{(0,2)}(Z)$, which are obtained in the Step 1, into Eqs. (\ref{a2deff}) and (\ref{reff}), and derive $a_{\rm 2D}$ and $R_s$.

In the following subsections we derive the  equations (\ref{ne1}, \ref{ne2}, \ref{a2deff},
\ref{reff}) and prove that $a_{\rm 2D}$ and $R_s$ can be calculated via the above steps.

\subsection{Lippmann-Schwinger Equation }

We consider the scattering of two equal-mass atoms in the quasi-2D setup, {\it i.e.}, the system described in Sec.~\ref{sec:level2}. The incident wave function $\Psi_{\bm k}^{\rm(in)}({\bm \rho},z_1,z_2)$ of the scattering is given in Eq.~(\ref{incident}), with ${\bm k}$ being the incident momentum. Since the inter-atomic interaction is described by the zero-range Huang-Yang pseudo potential $U_{\rm HY}(\bm{r})$ of Eq.~(\ref{uhy}), in our system the scattering only occurs in the subspace with $L_z=0$, where $L_z$ is the $z$-component of the angular momentum of the relative motion. Therefore, in the following we  focus on the projection of the scattering wave function  $\Psi_{\bm k}^{(+)}({\bm \rho},z_1,z_2)$ in this subspace, which is denoted as $\Psi_k^{(s)}\left(\rho, z_1, z_2\right)$, with $k=|{\bm k}|$ and $\rho=|{\bm \rho}|$. Notice that $\Psi_k^{(s)}$ is independent of the directions of ${\bm k}$ and  ${\bm \rho}$.
Furthermore, it is clear that $\Psi_k^{(s)}\left(\rho, z_1, z_2\right)$ can be expanded as
\begin{eqnarray}
\Psi_k^{(s)}(\rho, z_1, z_2)= \sum_{m, n=1}^{\infty}\phi_m(z_1) \phi_n\left(z_2\right)\Psi_k^{(m,n)}(\rho),\label{psiexp}
\end{eqnarray}
and satisfies the
the Schr\"odinger equation
\begin{eqnarray}
{\hat H}\Psi_k^{(s)}\left(\rho, z_1, z_2\right)=E\Psi_k^{(s)}\left(\rho, z_1, z_2\right),\ \ \ {\rm with}\ \
E=2E_1+\frac{k^2}{2\mu}<E_1+E_2,
\end{eqnarray}
as well as the  condition
\begin{eqnarray}
\lim_{\rho\rightarrow\infty}\Psi_k^{(m,n)}(\rho)&=&0,\ \ \ \ \ \ \ \ \ \ \ \ \ \ \  \ \ \ \ \ \ \ \ \ \  \ \ \ \ \ \ \ \ \ \ \ \ \ \ \  \ \ \ \ \ \ \ \ \ \  \ \ \ \ \ \ \ \  \ \ {\rm for} \ \ (m,n)\neq (1,1),\\
\Psi_k^{(1,1)}(\rho)&=&J_0(k\rho) + \frac{-1}{i\frac{\pi}{2}-\gamma- \ln\frac{ka_{\rm{2D}}}{2}-\frac{1}{2}R_{s}k^2}K_0(-ik\rho), \ \ \ \   {\rm for} \ \rho>0,
\label{conapp1}
\end{eqnarray}
where
$J_0$ and $K_0$ are the Bessel function of the first kind and the modified Bessel function of the second kind, respectively.
Notice that Eq.~(\ref{conapp1}) is correct in all the regions with $\rho>0$, due to the fact that $U_{\rm HY}(\bm{r})$ is a zero-range potential.

For the convenience of  calculation, here we introduce  another wave function
\begin{eqnarray}
\Phi_k(\rho, z_1, z_2)\equiv \sum_{m, n=1}^{\infty}\phi_m(z_1) \phi_n\left(z_2\right)\Phi_k^{(m,n)}(\rho),\label{phiexp}
\end{eqnarray}
which satisfies the Schr\"odinger equation
\begin{eqnarray}
{\hat H}\Phi_k\left(\rho, z_1, z_2\right)=E\Phi_k\left(\rho, z_1, z_2\right),
\end{eqnarray}
and the long-range boundary condition
\begin{eqnarray}
\lim_{\rho\rightarrow\infty}\Phi_k^{(m,n)}(\rho)&=&0,\ \ \ \ \ \ \ \ \ \ \ \ \ \ \  \ \ \ \ \ \ \ \ \ \  \   {\rm for} \ \ (m,n)\neq (1,1),\\
\Phi_k^{(1,1)}(\rho)&=&J_0(k\rho)+{\cal C}X_0(k,\rho),\ \ \ {\rm for}\ \rho>0. \label{conapp}
\end{eqnarray}
Here  the function $X_0(k,\rho)$ is defined as
\begin{eqnarray}
X_0(k, \rho)= K_0(-ik\rho) + \left[-\frac{i\pi}{2}+\gamma+\ln(k\ell_z)\right] J_0(k\rho).
\end{eqnarray}
and ${\cal C}$ is a to-be-determined constant. The explicit meaning of the condition (\ref{conapp}) is: If we express $\lim_{\rho\rightarrow\infty}\Phi_k^{(1,1)}(\rho)$ as the linear  combination of the functions $J_0(k\rho)$ and $X_0(k, \rho)$, then the  coefficient of $J_0(k\rho)$ is 1.

The function $\Phi_k\left(\rho, z_1, z_2\right)$ and the  scattering wave function   $\Psi_k^{(s)}\left(\rho, z_1, z_2\right)$ with $L_z=0$ satisfy the same Schr\"odinger equation and the same
long-range boundary condition for the excited transverse  modes,
but different long-range boundary condition for the ground transverse  mode. This fact yields that $\Phi_k\left(\rho, z_1, z_2\right)$  is proportional to $\Psi_k^{(s)}\left(\rho, z_1, z_2\right)$, {\it i.e.},
\begin{eqnarray}
\Phi_k\left(\rho, z_1, z_2\right) \propto \Psi_k^{(s)}\left(\rho, z_1, z_2\right).\label{prop}
\end{eqnarray}

In this and the next subsection, we investigate the Lippmann-Schwinger equation (LSE) of  $\Phi_k\left(\rho, z_1, z_2\right)$, and introduce the functions $\eta^{(0,2)}(Z')$ from this LSE.
In Appendix\ref{sss} we  derive $a_{\rm 2D}$ and $R_s$ from $\Phi_k\left(\rho, z_1, z_2\right)$.

The LSE of  $\Phi_k\left(\rho, z_1, z_2\right)$ can be expressed as
 ($\hbar=1$):
\begin{align}
  \Phi_k\left(\rho, z_1, z_2\right)=\phi_1(z_1) \phi_1(z_2)J_0(k\rho)+ \int_{-\infty}^{+\infty}dz_1'\int_{-\infty}^{+\infty}dz_2'\int_0^{\infty}d\rho^\prime
  G_E\bigg(\rho, z_1, z_2 ; \rho^\prime, z_1^{\prime}, z_2^{\prime}\bigg) D_k\left(\rho^\prime, z_1^{\prime}, z_2^{\prime}\right).\label{eq.LS-equation}
\end{align}
and the function $D_k\left(\rho^\prime, {z_1^\prime, z_2^\prime}\right)$ is defined as
\begin{equation}
 D_k\left(\rho^\prime, z_1^\prime,z_2^\prime\right) ={\frac{2 a_{\rm{3D}}}{\mu}}
 \delta(z_1^\prime-z_2^\prime)\delta(\rho^\prime)
  \frac{1}{\rho^\prime}\frac{\partial}{\partial \rho^\prime}\bigg[\rho^\prime\Phi_k\big({\rho^\prime},z_1^\prime,z_2^\prime\big)\bigg].
\end{equation}
Additionally, in Eq.~(\ref{eq.LS-equation})
$G_E$ is the free Green's function  in the subspace $L_z=0$, and can be expressed as:
\begin{eqnarray}
  G_E\bigg(\rho, z_1, z_2 ; \rho^\prime, z_1^{\prime}, z_2^{\prime}\bigg)= \sum_{m, n=1}^{\infty}\phi_m(z_1) \phi_m\left(z_1^{\prime}\right)\phi_n(z_2)\phi_n\left(z_2^{\prime}\right) g_{2D}^{(E-E_m-E_n)}({\rho},{\rho}^{\prime}),\label{eq.G4d}
\end{eqnarray}
with
\begin{eqnarray}
 g_{2D}^{({\cal E})}(\rho,\rho') &=&
 \left\{
    \begin{array}{cc}
      \tilde{\alpha}(\xi,\rho') J_0(\xi\rho) & (\rho<\rho') \\
      \\
      \tilde{\beta}(\xi,\rho') X_0(\xi,\rho) & (\rho>\rho')
    \end{array}\right., \hspace{1cm} ({\rm for}\ {\cal E}\geq 0);\\
    \nonumber\\
        \nonumber\\
    g_{2D}^{({\cal E})}(\rho,\rho') &=&
    \left\{
    \begin{array}{cc}
      \alpha(\kappa,\rho') I_0(\kappa\rho) & (\rho<\rho')\\
      \\
      \beta(\kappa,\rho') K_0(\kappa\rho) & (\rho>\rho')
    \end{array}\right., \hspace{1.2cm} ({\rm for}\ {\cal E}< 0).
\end{eqnarray}
Here
\begin{eqnarray}
\xi=\sqrt{2\mu{\cal E}},\ \ \ \ \kappa=\sqrt{2\mu|{\cal E}|},
\end{eqnarray}
$I_j$ and $K_j$  ($j=0,1,2,...$) are the first and second kinds of modified Bessel function, respectively,
and $\tilde{\alpha}$, $\tilde{\beta}$, $\alpha$ and
$\beta$ are defined as
\begin{eqnarray}
  \tilde{\alpha}(\xi,\rho') &=&\frac{2\mu X_0(\xi,\rho')}{J_0(\xi\rho')X_0^\prime(\xi,\rho')-J_0^\prime(\xi\rho')X_0(\xi,\rho')},\ \ \
  \tilde{\beta}(\xi,\rho') =\frac{2\mu J_0(\xi\rho')}{J_0(\xi\rho')X_0^\prime(\xi,\rho')-J_0^\prime(\xi\rho')X_0(\xi,\rho')},\\
   \alpha(\kappa,\rho') &=& \frac{2\mu K_0(\kappa\rho')}{\kappa\left[K_0(\kappa\rho')I_1(\kappa\rho')-K_1(\kappa\rho')I_0(\kappa\rho')\right]},\
    \beta(\kappa,\rho') = \frac{2\mu I_0(\kappa\rho')}{\kappa\left[K_0(\kappa\rho')I_1(\kappa\rho')-K_1(\kappa\rho')I_0(\kappa\rho')\right]},
\end{eqnarray}
with
$J_0^\prime(\xi\rho')=\frac{d }{d\rho'}J_0^\prime(\xi\rho')$ and $X_0^\prime(\xi\rho')=\frac{d }{d\rho'}X_0^\prime(\xi\rho')$.

\subsection{Definition of the Functions  $\eta^{(0,2)}(Z')$}

Now we re-express Eq.~(\ref{eq.LS-equation})  as
\begin{eqnarray}
  \Phi_k(\rho, z_1, z_2)=\phi_1(z_1)\phi_1(z_2)J_0(k\rho)+ \frac{1}{2}\int_{-\infty}^{+\infty}\mathrm{d}Z'\Omega_k\left(\rho,z_1, z_2,Z'\right)\eta_k(Z'),\label{lser}
\end{eqnarray}
%
%
where the functions $\eta_k(Z)$ and $\Omega_k(\rho,z_1, z_2,Z') $ are given by
\begin{equation}
  \eta_k(Z)=\ \lim_{\rho\to 0}\frac{2a_{\rm{3D}}}{\mu}\frac{\partial}{\partial \rho}\left[\rho\cdot\Phi_k(\rho,Z,Z)\right],\label{bp1}
\end{equation}
and
\begin{eqnarray}
  \Omega_k(\rho,z_1,z_2,Z')
  =
  \sum_{m,n=1}^{\infty}\phi_m(z_1)\phi_n(z_2)\phi_m(Z^{\prime})\phi_n(Z^{\prime})\cdot\Lambda_{\rm{2D}}^{(E-E_m-E_n)}(\rho),\label{eq.omega}
\end{eqnarray}
respectively. Here the function $\Lambda_{\rm{2D}}^{({\cal E})}(\rho)$ is defined as
  \begin{eqnarray}
    \Lambda_{\rm{2D}}^{({\cal E})}(\rho) = \left\{
    \begin{array}{ll}
      -2\mu X_0(\sqrt{2\mu{\cal E}}, \rho),
      &\hspace{1cm} ({\cal E}\geq 0) \\
      \\
      -2\mu K_0(\sqrt{2\mu|{\cal E}|}\rho), &\hspace{1cm}({\cal E}< 0)\
    \end{array}\right..\label{eq.x0}
  \end{eqnarray}


In the short-range regime $\rho\ll 1/k$, we can expand the scattering wave function $\Phi_k(\rho, z_1, z_2)$ and the function $\eta_k(Z') $ as
\begin{eqnarray}
\Phi_k(\rho, z_1, z_2) = \Phi^{(0)}(\rho, z_1, z_2) + k^2 \Phi^{(2)}(\rho, z_1, z_2)+ \mathcal{O}(k^4),\label{psiexp}
\end{eqnarray}
and
\begin{eqnarray}
    \eta_k(Z') =& \eta^{(0)}(Z') + k^2\eta^{(2)}(Z') + \mathcal{O}(k^4),
\end{eqnarray}
respectively, with $\eta^{(j)}(Z')$ $(j=0,2)$ being related to $\Phi^{(j)}(\rho, z_1,z_2)$ via
\begin{eqnarray}
  \eta^{(j)}(Z')=\ \frac{2a_{\rm{3D}}}{\mu}\left.\frac{\partial}{\partial \rho'}\bigg[\rho'\cdot\Phi^{(j)}(\rho',Z',Z')\bigg]\right|_{\rho'=0},\hspace{1cm} (j=0,2).
  \label{eq.eta}
\end{eqnarray}
Moreover, by expanding both sides of Eq.~(\ref{lser})  in the region with $\rho\ll 1/k$,
we obtain
\begin{eqnarray}
  \Phi^{(0)}(\rho, z_1, z_2)&=&\phi_1(z_1)\phi_1(z_2)+ \frac{1}{2}\int_{-\infty}^{+\infty}\Omega_{k=0}(\rho,z_1,z_2,Z')\eta^{(0)}(Z')\mathrm{d}Z',\label{eq.LS-omega0}\\
  \nonumber\\
  \Phi^{(2)}(\rho, z_1, z_2)&=&-\phi_1(z_1)\phi_1(z_2)\frac{\rho^2}{4}+ \frac{1}{2}\int_{-\infty}^{+\infty}\Omega^{(2)}(\rho,z_1,z_2,Z')\eta^{(0)}(Z')\mathrm{d}Z'+ \frac{1}{2}\int_{-\infty}^{+\infty}\Omega_{k=0}(\rho,z_1,z_2,Z')\eta^{(2)}(Z')\mathrm{d}Z',\nonumber\\
  \label{eq.LS-omega2}
\end{eqnarray}
where the function $\Omega^{(2)}(\rho,z_1,z_2,Z')$ is defined as
\begin{eqnarray}
\Omega^{(2)}(\rho,z_1,z_2,Z') &=&\sum_{m,n=1}^{\infty}\phi_m(z_1)\phi_n(z_2)\phi_m(Z')\phi_n(Z')\Lambda_{2}^{(mn)}(\rho),\label{omega2}
\end{eqnarray}
with
\begin{eqnarray}
  \Lambda_2^{(mn)}(\rho) =
  \left\{
  \begin{array}{ll}
    -\dfrac{\mu}{2}\rho^2\ln\left(\dfrac{\rho}{2\ell_z}\right), & (m=n=1)\\
    \\
    \sqrt{\dfrac{2\mu}{|E_m+E_n|}}\rho K_1\left[\sqrt{2\mu(E_m+E_n)}\rho\right], & ((m,n)\neq (1,1))
  \end{array}\right..
\end{eqnarray}

\subsection{Derivation of Eqs.~(\ref{a2deff}, \ref{reff})}
\label{sss}

Substituting Eqs.~(\ref{eq.LS-omega0}, \ref{eq.LS-omega2}) into Eq.~(\ref{psiexp}) and then into Eq.~(\ref{phiexp}), we find that  the component $\Phi_k^{(1,1)}(\rho)$
in the expansion (\ref{phiexp}) of $\Phi_k(\rho, z_1, z_2)$
has the following short-range behavior:
\begin{eqnarray}
\Phi_k^{(1,1)}(\rho)=1 + \mu\ln\left(\frac{\rho}{2\ell_z}\right)\left\{\int_{-\infty}^{+\infty}\mathrm{d}Z'
\phi_1(Z')^{ 2}
\left[\eta^{(0)}(Z')+k^2\eta^{(2)}(Z')\right]\right\}+\mathcal{O}(\rho)+\mathcal{O}(k^4),
\ \ \ \ {\rm for}\ \rho\ll 1/k.
\label{psik11}
\end{eqnarray}

On the other hand, the property (\ref{conapp1}) of the component $\Psi_k^{(1,1)}(\rho)$ of the scattering wave function $\Psi_k^{(s)}\left(\rho, z_1, z_2\right)$  with $L_z=0$ satisfies
\begin{eqnarray}
  \Psi_k^{(1,1)}\left(\rho, z_1, z_2\right) \propto \ln(\rho/a_{\rm{2D}}) -\frac{1}{2}R_sk^2 +\mathcal{O}(\rho)+\mathcal{O}(k^4),\ \ \ \ {\rm for}\ \rho\ll 1/k.\label{eq.pure-2d-asym}
\end{eqnarray}
Furthermore, as shown in Eq.~(\ref{prop}), we have $\Phi_k\left(\rho, z_1, z_2\right) \propto \Psi_k^{(s)}\left(\rho, z_1, z_2\right)$, which yields $\Phi_k^{(1,1)}(\rho)\propto\Psi_k^{(1,1)}\left(\rho, z_1, z_2\right)$. Combining this fact and Eqs.~(\ref{psik11}, \ref{eq.pure-2d-asym}), we obtain
\begin{eqnarray}
&&1 + \mu\ln\left(\frac{\rho}{2\ell_z}\right)\left\{\int_{-\infty}^{+\infty}\mathrm{d}Z'
\phi_1(Z')^{ 2}
\left[\eta^{(0)}(Z')+k^2\eta^{(2)}(Z')\right]\right\}\nonumber\\
&=&
\mu\left\{\int_{-\infty}^{+\infty}\mathrm{d}Z'
\phi_1(Z')^{ 2}
\left[\eta^{(0)}(Z')+k^2\eta^{(2)}(Z')\right]\right\}\left[
\ln(\rho/a_{\rm{2D}}) -\frac{1}{2}R_sk^2\right],
\end{eqnarray}
which gives
\begin{eqnarray}
 a_{\rm{2D}}&=&\ 2\ell_z\exp\left[\frac{-1}{\mu \int\phi_1(Z')^2\eta^{(0)}(Z')\mathrm{d}Z'}\right],\label{a2deff}\\
  \nonumber\\
  R_{s}&=&\frac{2}{\mu}\frac{\int\phi_1(Z')^2\eta^{(2)}(Z')\mathrm{d}Z'}{\left[ \int\phi_1(Z')^2\eta^{(0)}(Z')\mathrm{d}Z'\right]^2}.\label{reff}
\end{eqnarray}



\subsection{Derivations of Eqs.~(\ref{ne1}, \ref{ne2})}
\label{sec.app-b}


By combining Eq.~(\ref{bp1}) and the Bethe-Peierls boundary condition satisfied by the wave function $\Phi_k(\rho,z_1,z_2)$, which is due to the Huang-Yang pseudo potential used in our calculation, we obtain
\begin{eqnarray}
  \Phi_k(\rho,Z,Z)&=&-\frac{\mu}{2}\eta(Z)\left(\frac{1}{\rho} -\frac{1}{a_{\rm{3D}}} \right) + \mathcal{O}(\rho).
\end{eqnarray}
This result and Eq.~(\ref{eq.eta}) further yield that
\begin{eqnarray}
  \Psi^{(j)}(\rho,Z,Z)&=&-\frac{\mu}{2}\eta^{(j)}(Z)\left(\frac{1}{\rho} -\frac{1}{a_{\rm{3D}}} \right) + \mathcal{O}(\rho),\ \ \ \ (j=0,2).\label{eqeta2}
\end{eqnarray}
Using Eq.~(\ref{eqeta2}), we can formally express the functions $\eta^{(0,2)}(Z)$ as
\begin{eqnarray}
  \eta^{(j)}(Z)=\frac{2a_{\rm{3D}}}{\mu}\hat{\mathbb O}^{(j)}(Z)\bigg[\Psi^{(j)}(\rho,Z,Z)\bigg],\ \ \ \ (j=0,2).\label{eq.eqhatO}
\end{eqnarray}
Here  $\hat{\mathbb O}^{(j)}(Z)$ is defined as
\begin{eqnarray}
 \hat{\mathbb O}^{(j)}(Z)\bigg[\Psi^{(j)}(\rho,Z,Z)\bigg]\equiv\lim_{\rho\to 0}\left[\Psi^{(j)}(\rho,Z,Z) + \frac{\mu}{2\rho}\eta^{(j)}(Z)\right],\ \ \ \ (j=0,2).
\end{eqnarray}
Subsequently, by taking $z_1=z_2=Z$ and acting $\hat{\mathbb O}(Z)$ on both sides of Eqs.(\ref{eq.LS-omega0}, \ref{eq.LS-omega2}), we obtain the integral equation for $\eta^{(0,2)}(Z)$ as
\begin{eqnarray}
  \frac{\mu}{2a_{\rm{3D}}}\eta^{(0)}(Z)&=&\phi_1^2(Z)+ \frac{1}{2}\lim_{\rho\to 0}\mathcal{I}^{(0)}(\rho,Z)\label{eq.eta0}\\
  \nonumber\\
  \frac{\mu}{2a_{\rm{3D}}}\eta^{(2)}(Z)&=& \frac{1}{2}\int_{-\infty}^{+\infty}\Omega^{(2)}(0,Z,Z,Z')\eta^{(0)}(Z')\mathrm{d}Z'+ \frac{1}{2}\lim_{\rho\to 0}\mathcal{I}^{(2)}(\rho,Z),\label{eq.eta2}
\end{eqnarray}
where the functions $\mathcal{I}^{(0,2)}(\rho,Z)$ are defined as:
\begin{eqnarray}
  \mathcal{I}^{(j)}(\rho,Z)&\equiv&\int_{-\infty}^{+\infty}\Omega_{k=0}(\rho,Z,Z,Z')\eta^{(j)}(Z')\mathrm{d}Z' + \frac{\mu}{\rho}\eta^{(j)}(Z),\ \ \ \ (j=0,2).\label{eq.I}
\end{eqnarray}

We can furthere re-express the terms $\lim_{\rho\to 0}\mathcal{I}^{(0,2)}(\rho,Z)$ of Eqs. (\ref{eq.eta0}, \ref{eq.eta2}) as Hadamard finite part integral. Explicitly,
 using the fact
\begin{eqnarray}
  \frac{1}{\rho} &=& \frac{1}{\pi}\int_{-\infty}^{+\infty}\frac{b}{\rho^2+b^2(z-z^{\prime})^2}\mathrm{d}z^{\prime}, \ \ \ \forall b>0,
      \end{eqnarray}
we can obtain
\begin{eqnarray}
   \lim_{\rho\to 0}\mathcal{I}^{(j)}(\rho,Z)&=&\lim_{\rho\to 0}\left[\int_{-\infty}^{+\infty}\Omega_{k=0}(\rho,Z,Z,Z')\eta^{(j)}(Z')\mathrm{d}Z' + \int_{-\infty}^{+\infty}\frac{\mu}{\pi}\frac{b\eta^{(j)}(Z)}{\rho^2+b^2(Z-Z^{\prime})^2}\mathrm{d}Z'\right]\nonumber\\
  &=&\lim_{\rho\to 0}\lim_{\epsilon\to 0}\left[\int_{-\infty}^{Z'-\epsilon}\Omega_{k=0}(\rho,Z,Z,Z')\eta^{(j)}(Z')\mathrm{d}Z'+\int^{+\infty}_{Z^{\prime}+\epsilon}\Omega_{k=0}(\rho,Z,Z,Z')\eta^{(j)}(Z')\mathrm{d}Z'\right.\nonumber\\
  &&\ \ \ \ \ \ \ \ \ \ \left.+ \int_{-\infty}^{Z'-\epsilon}\frac{\mu}{\pi}\frac{b\eta^{(j)}(Z)}{\rho^2+b^2(Z-Z^{\prime})^2}\mathrm{d}Z' + \int_{Z'+\epsilon}^{+\infty}\frac{\mu}{\pi}\frac{b\eta^{(j)}(Z)}{\rho^2+b^2(Z-Z^{\prime})^2}\mathrm{d}Z'\right]. \label{ehad1}
    \end{eqnarray}
Notice that Eq.~(\ref{ehad1}) is correct for arbitrary positive $b$.
Here we define $b_0$ to be a special positive number, which satisfies the following condition:
When $b=b_0$ the integration in Eq.~(\ref{ehad1}) uniformly  convergences in the limit $\rho\rightarrow 0$, so that the orders of $\lim_{\rho\rightarrow 0}$ and
performing the integration can be exchanged. Later we will determine the value of $b_0$ according to the behavior of $\Omega_{k=0}(\rho,Z,Z,Z')$ in the limits $\rho\to 0$.

Taking $b=b_0$, we can re-express Eq.~(\ref{ehad1}) as
\begin{eqnarray}
  \lim_{\rho\to 0}\mathcal{I}^{(j)}(\rho,Z)  &=& \mathbb{Z}\int \Omega_{k=0}(\rho,Z,Z,Z')\eta^{(j)}(Z')\mathrm{d}Z',\nonumber\\
  &\equiv&\lim_{\epsilon\to 0}\left[\int_{-\infty}^{Z'-\epsilon} \Omega_{k=0}(0,Z,Z,Z')\eta^{(j)}(Z')\mathrm{d}Z' + \int_{Z'+\epsilon}^{+\infty}\Omega_{k=0}(0,Z,Z,Z')\eta^{(j)}(Z')\mathrm{d}Z' + \frac{2}{b_0}\frac{\mu}{\pi}\frac{\eta^{(j)}(Z)}{\epsilon} \right],\nonumber\\
  &&\hspace{12cm} (j=0,2).
\label{ehad2}
\end{eqnarray}
where $\mathbb{Z}\int\cdots\mathrm{d}Z^{\prime}$ refers to Hadamard finite part integral. Notice that the Hadamard finite part integral of Eq.~(\ref{ehad2}) is only determined by the values of $\Omega_{k=0}(0,Z,Z,Z')$ for $Z\neq Z'$.

Now we further re-express the function $\Omega_{k=0}(0,Z,Z,Z')$ and derive the value of $b_0$. Using the relation{
\begin{eqnarray}
  -2\mu K_0\left(\sqrt{2\mu|\mathcal{E}|}\rho\right)&=&\int_0^{+\infty}{\mathrm d}\beta \frac{\mu}{\beta}e^{\beta\mathcal{E}-\frac{\mu\rho^2}{2\beta}},\ \ \ \ \ \ \ (\mathcal{E}<0),\label{eq.laplace-k}
\end{eqnarray}}
we find that the function $\Omega_{k=0}(\rho,Z,Z,Z')$ defined in Eq.~(\ref{eq.omega}) can be re-expressed as
\begin{eqnarray}
  \Omega_{k=0}(\rho,Z,Z,Z')&=&-2\mu X_0(k,\rho)\phi_1^2(Z)\phi_1^2(Z')
  + 2\pi\sum_{(m,n)\neq(1,1)}{\int_0^{+\infty}{\mathrm d}\tau \frac{\mu}{\tau}e^{\left(2E_1-E_n-E_m\right)\tau-\frac{\mu\rho^2}{2\tau}}\phi_n(Z) \phi_n(Z')\phi_m(Z)\phi_m(Z')}
  \label{eq.new-omega1}\nonumber\\
  &=&-2\mu X_0(k,\rho)\phi_1^2(Z)\phi_1^2(Z')\!
  -\!\int_0^{+\infty}\!\mathrm{d}\tau\frac{\mu}{\tau}e^{2E_1\tau-\frac{\mu\rho^2}{2\tau}}\left[g_{\mathrm{1D}}^2\left(Z,Z',\tau\right)-\!e^{-2 E_1 \tau}\phi_1^2(Z)\phi_1^2\left(Z'\right)\right],\nonumber\\
  \label{eq.new-omega2}
\end{eqnarray}
where $g_{\mathrm{1D}}\left(Z,Z',\tau\right)$ is the imaginary propagator of the 1D confinement Hamiltonian:
\begin{eqnarray}
  g_{\mathrm{1D}}\left(Z,Z',\tau\right)
  &=&{\sum_{n=1}^{+\infty}\phi_n(Z)\phi_n(Z')\exp(-E_n\tau)}.\label{eq.imag-g1d}
\end{eqnarray}
For the box confinement ($\lambda=\infty$), we have
\begin{eqnarray}
  g_{\mathrm{1D}}\left(Z,Z',\tau\right)
  &=&\frac{1}{4\ell_z}\left\{\vartheta_3\left[\frac{\pi(Z-Z')}{4\ell_z}, e^{-E_1\tau}\right]-\vartheta_4\left[\frac{\pi(Z-Z')}{4\ell_z}, e^{-E_1\tau}\right]\right\},\label{g1d}
\end{eqnarray}
where $\vartheta_{n}(u,q)$ are elliptic theta functions. To calculate $\Omega_{k=0}(0,Z,Z,Z')$, we further re-write Eq.~(\ref{eq.new-omega2}) as
\begin{eqnarray}
  \Omega_{k=0}(\rho,Z,Z,Z')&=&\phi_1^2(Z)\phi_1^2(Z')2\mu\left[K_0(\kappa_c\rho)-X_0(k,\rho)\right] -\int_0^{+\infty}\mathrm{d}\tau\frac{\mu}{\tau}e^{2E_1\tau-\frac{\mu\rho^2}{2\tau}}\left[g_{\mathrm{1D}}^2\left(Z,Z',\tau\right)\right.\nonumber\\
  &&\left.-\left(e^{-2 E_1 \tau}-e^{-E_{\mathrm{c}}\tau}\right)\phi_1^2(Z)\phi_1^2\left(Z'\right)\right],\label{eq.new-omega3}
\end{eqnarray}
where  $E_c$ is an arbitrary energy satisfying $E_c>2E_1$, and $\kappa_c=\sqrt{2\mu(E_c-2E_1)}$. Eq.~(\ref{eq.new-omega3}) yields that
\begin{eqnarray}
  \Omega_{k=0}(0,Z,Z,Z')&=&-2\mu\ln(\kappa_c\ell_z e^{\gamma}) \phi_1^2(Z) \phi_1^2\left(Z'\right) -\int_0^{+\infty}\mathrm{d}\tau\frac{\mu}{\tau}\left[e^{2E_1\tau}g_{\mathrm{1D}}^2\left(Z, Z', \tau\right) - \left(1-e^{-\kappa_c^2\tau}\right) \phi_1^2(Z) \phi_1^2\left(Z'\right)\right].\nonumber\\
  \label{eq.new-omega4}
\end{eqnarray}
Notice that the integral in Eq.~(\ref{eq.new-omega4}) converges.
Similarly, $\Omega^{(2)}(0,Z,Z,Z')$ is obtained as
\begin{eqnarray}
  \Omega^{(2)}(0,Z,Z,Z')= -\int_0^{+\infty}\mathrm{d}\tau\mu\left[e^{2E_1\tau}g_{\mathrm{1D}}^2\left(Z, Z', \tau\right) - \phi_1^2(Z) \phi_1^2\left(Z'\right)\right].\label{eo2}
\end{eqnarray}

On the other hand, Due to the fact that in the $\tau\to 0$ limit for any form of $V_{\mathrm{conf}}(z)$
\begin{eqnarray}
  g_{\mathrm{1D}}\left(Z,Z',\tau\right)&\simeq&\sqrt{\frac{m}{2\pi\tau}}e^{-\frac{m(Z-Z^{\prime})^2}{2\tau}},
\end{eqnarray}
the asymptotic behavior of $\Omega_{k=0}(\rho,Z,Z,Z')$ in the limit $\rho\to 0$ and $Z-Z^{\prime}\to 0$ is obtained by
\begin{eqnarray}
  \Omega_{k=0}(\rho,Z,Z,Z')&\simeq&-\frac{\mu}{\pi}\frac{2}{\rho^2+4(Z-Z^{\prime})^2}.
\end{eqnarray}
which gives
\begin{eqnarray}
b_0=2.
\end{eqnarray}

In conclusion, Eqs.~(\ref{eq.eta}-\ref{eq.LS-omega2}\label{sec.app-b}) are reduced to
\begin{eqnarray}
  \frac{\mu}{2 a_{\rm{3D}}}\eta^{(0)}(Z)&=&\phi_1^2(Z) + \frac{1}{2}\mathbb{Z}\int \Omega_{k=0}(0,Z,Z,Z') \eta^{(0)}\left(Z'\right)\mathrm{d}Z', \label{ne1} \\
  \frac{\mu}{2 a_{\rm{3D}}}\eta^{(2)}(Z)&=&\frac{1}{2}\int \Omega^{(2)}(0,Z,Z,Z') \eta^{(0)}\left(Z'\right)\mathrm{d}Z' + \frac{1}{2}\mathbb{Z}\int \Omega_{k=0}(0,Z,Z,Z') \eta^{(2)}\left(Z'\right)\mathrm{d}Z',
  \label{ne2}
\end{eqnarray}
with the Hadamard integral $\mathbb Z$ being defined in Eq.~(\ref{ehad2}) with $b_0=2$, and the functions $\Omega_{k=0}(0,Z,Z,Z')$ and $\Omega^{(2)}(0,Z,Z,Z')$ being given in Eq.~(\ref{eq.new-omega4}) and Eq.~(\ref{eo2}), respectively. We derive $\eta^{(0,2)}(Z)$ by solving Eqs.~(\ref{ne1}, \ref{ne2}) with standard numerical approaches.

\section{Calculations for the Bound States}

\subsection{Summary of the Main Steps}

Now we show our approach for the calculation of the binding energy $E_{\rm binding}$ and transverse-excited-state probability $P_{\rm ex}$ for the bound states, which is similar as the one in Appendix A for the scattering problem.

As in Appendix A, for the convenience of the readers  we first summarize  our approach in this subsection, and show the related proof in the Appendix A2 and A3.

Our approach includes the following two steps:

{\bf Step 1:} Numerically solving the homogeneous integral equation (\ref{eq.newetab3}),  and then deriving the bound-state energy $E_b$ as well as the function $\eta_b(Z)$ (up to a constant factor). The binding energy $E_{\rm binding}$ can be obtained   via the value of $E_b$ and the relation (\ref{ebb}) of our main text.

{\bf Step 2:} Substituting the function $\eta_b(Z)$ obtained in Step 1 into Eqs.~(\ref{pe1}, \ref{pe3}), and then obtaining the value of $P_{\rm ex}$.

In the following subsections we show the derivation of Eq.~(\ref{eq.newetab3}) and Eqs.~(\ref{pe1}, \ref{pe3}).

\subsection{Derivation of Eq.~(\ref{eq.newetab3})}

The Schr\"odinger equation for bound-state wave function $\Phi_b\left({\rho}, z_1, z_2\right) $ is
\begin{eqnarray}
  \hat{H}\Phi_b\left({\rho}, z_1, z_2\right) = E_b\Phi_b\left({\rho}, z_1, z_2\right),\label{sseb}
\end{eqnarray}
with the energy $E_b<2E_1$. Notice that the bound-state is in the subspace with $L_z=0$, and thus the wave function is independent of the direction of ${\bm \rho}$.
The  Schr\"odinger equation Eq.~(\ref{sseb}) can be re-expressed as a LSE-type integral equaiton:
\begin{align}
  \Phi_b\left(\rho, z_1, z_2\right) = \int_{-\infty}^{+\infty}dz_1'\int_{-\infty}^{+\infty}dz_2'\int_0^{\infty}d\rho^\prime
  G_{E_b}\bigg(\rho, z_1, z_2 ; \rho^\prime, z_1^{\prime}, z_2^{\prime}\bigg) D_b\left(\rho^\prime, z_1^{\prime}, z_2^{\prime}\right).\label{eq.lseb-0}
\end{align}
where the function $G_{E}$ is defined in  Eq.~(\ref{eq.G4d}), and
\begin{equation}
 D_b\left(\rho^\prime, z_1^\prime,z_2^\prime\right) =\frac{2 a_{\rm{3D}}}{\mu}
 \delta(z_1^\prime-z_2^\prime)\delta(\rho^\prime)
  \frac{1}{\rho^\prime}\frac{\partial}{\partial \rho^\prime}\bigg[\rho^\prime\Phi_b\big({\rho^\prime},z_1^\prime,z_2^\prime\big)\bigg].
\end{equation}
After integrating out the delta function in $D_b\left(\rho^\prime, z_1^\prime,z_2^\prime\right)$, we can re-express Eq.~(\ref{eq.lseb-0}) as
\begin{eqnarray}
  \Phi_b\left(\rho, z_1, z_2\right)&=&\frac{1}{2}\int \Omega_b\left(\rho',z_1,z_2,Z'\right) \eta_b\left(Z'\right)\mathrm{d}Z',\label{eq.lseb}
\end{eqnarray}
where the function $  \eta_b(Z)$ is defined as
\begin{equation}
  \eta_b(Z')=\ \left.\frac{2a_{\rm{3D}}}{\mu}\frac{\partial}{\partial \rho'}\bigg[\rho'\cdot\Phi_b(\rho',Z',Z')\bigg]\right|_{\rho'=0}, \label{eq.etab}
\end{equation}
and $\Omega_b(\rho,z_1,z_2,Z')$ is specifically given by
\begin{eqnarray}
  \Omega_b(\rho,z_1,z_2,Z')
  =
  -\sum_{m,n=1}^{\infty}\phi_m(z_1)\phi_n(z_2)\phi_m(Z^{\prime})\phi_n(Z^{\prime})\cdot2\mu K_0\left(\kappa_{m,n}\rho\right).\label{b6}
\end{eqnarray}
with $\kappa_{m,n}=\sqrt{-2\mu(E_b-E_m-E_n)}$.

Furthermore, using the method similar to the one in Appendix \ref{sec.app-b},
we can derive the homogeneous integral equation for $E_b$ and $\eta_b(Z)$:
\begin{eqnarray}
  \frac{\mu}{2 a_{\rm{3D}}}\eta_b(Z)&=&\frac{1}{2}\mathbb{Z}\int F_b\left(E_b,Z, Z'\right) \eta_b\left(Z'\right)\mathrm{d}Z',\label{eq.newetab3}
\end{eqnarray}
where Hadamard integral $\mathbb Z$ is defined in Eq.~(\ref{ehad2}) with $b_0=2$, and the function $F_b\left(E_b,Z,Z'\right)$ is defined as
\begin{eqnarray}
  F_b\left(E_b,Z,Z'\right)&=&-\int_0^{+\infty}\mathrm{d}\tau\frac{\mu}{\tau}e^{E_b\tau}g_{\mathrm{1D}}^2\left(Z, Z', \tau\right),
\end{eqnarray}
with the function $g_{\mathrm{1D}}$ being defined in Eq.~(\ref{g1d}).

\subsection{Derivation of Eqs.~(\ref{pe1}, \ref{pe3})}

Using Eq.~(\ref{eq.lseb}) and Eq.~(\ref{b6}),
we can re-express the normalized bound-state wave function $\Phi_b\left(\rho, z_1, z_2\right)$ as
\begin{eqnarray}
  \Phi_b\left(\rho, z_1, z_2\right)&=&\sum_{m,n=1}^{\infty}\phi_m(z_1)\phi_n(z_2)\phi_b^{m,n}(\rho),
\end{eqnarray}
where $\phi_b^{m,n}(\rho) = A \tilde{\phi}_b^{m,n}(\rho)$, with A being the normalization factor  and
\begin{eqnarray}
  \tilde{\phi}_b^{m,n}(\rho) = -\mu K_0\left(\kappa_{m,n}\rho\right)\int\phi_m(Z')\phi_n(Z')\eta_b(Z') \mathrm{d} Z'.
\end{eqnarray}
Therefore,
for the two atoms in the bound state,
the probability of
the transverse eigen-state
$\phi_n(z_1)\phi_m(z_2)$ ($m,n=1,2,3,...$) can be expressed as
\begin{eqnarray}
  P_{m,n} = 2\pi\int_0^{+\infty} \mathrm{d}{\rho} \vert\phi_b^{(m,n)}({\rho})\vert ^2.\label{pe1}
\end{eqnarray}
Thus, the transverse-excited-state probability $P_{\rm ex}$ of the bound state is given by
\begin{equation}
  P_{\rm ex} = 1-P_{1,1}.\label{pe3}
\end{equation}

\end{widetext}

\nocite{*}


\begin{thebibliography}{99}
\bibitem{ftn3}
The quantitative criterial of the condition (\ref{con}) depends on the specific system and the to-be-studied problem.
\bibitem{ftn1}
Our calculation method can be directly generalized to the cases of two atoms with un-equal masses.
\end{thebibliography}

\end{document}